\documentclass[aps,prd,preprint,tightenlines,
   showpacs,nofootinbib]{revtex4}
\newcommand{\PRE}[1]{{#1}} 

\usepackage{bm} \usepackage{epsfig}
\usepackage{multirow}

\newcommand{\ifb}{\text{fb}^{-1}}

\newcommand{\gev}{\text{GeV}}
\newcommand{\tev}{\text{TeV}}
\newcommand{\fb}{\text{fb}}

\newcommand{\eg}{{\em e.g.}}

\newcommand{\eqref}[1]{Eq.~(\ref{#1})}
\newcommand{\eqsref}[2]{Eqs.~(\ref{#1}) and (\ref{#2})}
\newcommand{\secref}[1]{Sec.~\ref{sec:#1}}

\newcommand{\figref}[1]{Fig.~\ref{fig:#1}}

\newcommand{\tableref}[1]{Table~\ref{table:#1}}

\newcommand{\bea}{\begin{eqnarray}}
\newcommand{\eea}{\end{eqnarray}}

\newcommand{\unp}{{\cal U}}

\begin{document}

\preprint{UCI-TR-2007-50}

\title{ \PRE{\vspace*{1.5in}}Unparticle Self-Interactions and
  Their Collider Implications \PRE{\vspace*{0.3in}} }

\author{Jonathan L.~Feng}
\affiliation{Department of Physics and Astronomy, University of
California, Irvine, CA 92697, USA \PRE{\vspace*{.5in}} }

\author{Arvind Rajaraman}
\affiliation{Department of Physics and Astronomy, University of
California, Irvine, CA 92697, USA \PRE{\vspace*{.5in}} }

\author{Huitzu Tu%
\PRE{\vspace*{.2in}} }
\affiliation{Department of Physics and Astronomy, University of
California, Irvine, CA 92697, USA \PRE{\vspace*{.5in}} }


\begin{abstract}
\PRE{\vspace*{.3in}} In unparticle physics, operators of the conformal
sector have self-interactions, and these are unsuppressed for strong
coupling.  The 3-point interactions are completely determined by
conformal symmetry, up to a constant.  We do not know of any
theoretical upper bounds on this constant.  Imposing current
experimental constraints, we find that these interactions mediate
spectacular collider signals, such as $pp \to \unp \to \unp \unp \to
\gamma \gamma\gamma\gamma$, $\gamma\gamma ZZ$, $ZZZZ$, $\gamma\gamma
l^+ l^-$, $ZZl^+ l^-$, and $4l$, with cross sections of picobarns or
larger at the Large Hadron Collider.  Self-interactions may therefore
provide the leading discovery prospects for unparticle physics.
\end{abstract}


\maketitle

\section{Introduction}
\label{sec:intro}

In unparticle physics, the standard model is extended by couplings to
a conformal sector through interactions of the form $O O_{\text{SM}}$,
where $O$ is an operator of the conformal sector, and $O_{\text{SM}}$
is a standard model operator~\cite{Georgi:2007ek}.  The conformal
sector may be weakly-coupled~\cite{Banks:1981nn} or
strongly-coupled~\cite{Intriligator:1995au}, but in all cases, the
conformality of the new sector leads to effects that cannot be
explained in terms of standard particle states.

To date, all unparticle studies are based on two key elements:~the
unparticle phase space, and the unparticle propagator.  Conformal
invariance fixes unparticle phase space~\cite{Georgi:2007ek}, which
enters processes with unparticles in the final state, such as $f
\bar{f} \to f \unp$.  Similarly, conformal invariance dictates the
form of the unparticle propagator~\cite{Georgi:2007si,Cheung:2007zza},
which determines virtual unparticle contributions to processes such as
$f \bar{f} \to \unp \to f \bar{f}$.  The forms of the unparticle phase
space and propagator imply that unparticles do not behave as standard
particles, but are more aptly interpreted as fractional numbers of
massless particles~\cite{Georgi:2007ek} or as collections of particles
with a particular distribution of masses~\cite{Stephanov:2007ry}.
These results are valid if the couplings of unparticles to the
standard model are all non-renormalizable.  If unparticle couplings
include the super-renormalizable operator $O h^2$, where $h$ is the
standard model Higgs boson, electroweak symmetry breaking breaks
conformal invariance~\cite{Fox:2007sy,Delgado:2007dx}.  This modifies
the unparticle propagator and implies that unparticle physics may be
probed only in a narrow conformal window, typically at energies
between 10 GeV and 1 TeV~\cite{Fox:2007sy,Bander:2007nd}. Such
energies are best probed at high energy colliders, and many studies
have investigated the collider implications of
unparticles~\cite{Georgi:2007ek,Georgi:2007si,%
Cheung:2007zza,Bander:2007nd,Greiner:2007hr,Kumar:2007af,%
colliderstudies}.  Models with conformal breaking may also share
characteristics with hidden valley
models~\cite{Strassler:2006im,Strassler:2008bv}.

Here we study a qualitatively new effect:~unparticle
self-interactions.  Such 3- and higher-point interactions are always
present in conformal theories, and mediate processes such as $g g \to
\unp \to \unp \cdots \unp$, with two or more unparticles in the final
state.  In the most interesting cases with strongly-coupled conformal
sectors, the creation of additional high $p_T$ unparticles in the
final state does not suppress the rate.  Multi-$\unp$ production
therefore differs from all known examples, such as $gg \to g \to g
\cdots g$ and $g g \to \gamma \cdots \gamma$, where the rate is
reduced with the addition of every high $p_T$ particle.

\begin{figure}[tb]
\begin{center}
\includegraphics*[width=9cm,clip=]{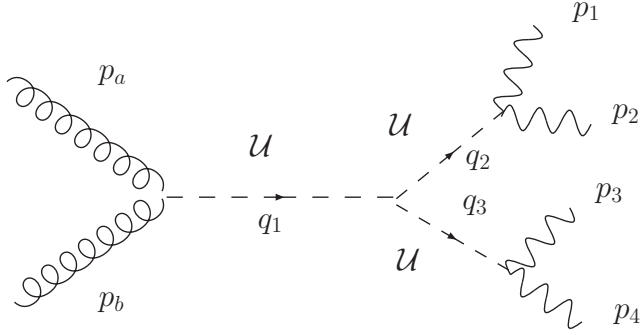}
\end{center}
\caption{The process $g g \to \gamma \gamma \gamma \gamma$ mediated by
  unparticle self-interactions.}
\label{fig:4gamma_feynmandiagram}
\end{figure}

In this paper, we focus on 3-point self-interactions.  These are the
natural starting point for several reasons.  First, unlike 4- and
higher-point self-interactions, 3-point interactions are completely
constrained by conformal invariance, up to a constant.  Second,
3-point interactions are the leading order at which induced signals
may be nearly background free.  For example, although the $\unp$
propagator induces signals like $p p \to \unp \to \gamma \gamma$,
3-point $\unp$ self-interactions mediate $pp \to \unp \to \unp \unp
\to \gamma \gamma\gamma\gamma, \gamma\gamma ZZ, ZZZZ, \gamma\gamma l^+
l^-, ZZl^+ l^-, 4l$, and many other spectacular signals through
subprocesses such as the one shown in \figref{4gamma_feynmandiagram}.
For this reason, the most promising signals for unparticle discovery
at colliders may in fact be those that are induced by 3-point
unparticle interactions.  (Note that, although the production of 3 or
more unparticles may also be unsuppressed, the requirement that they
convert back to standard model particles to be visible does, in fact,
imply that these are sub-dominant.)

In the following sections, we start from the 3-point correlation
function in position space and convert it to a form useful for Feynman
diagram calculations.  We then calculate collider rates, taking the
$4\gamma$ signal as our example.  Our final result is determined up to
the constant entering the 3-point correlation function.  As far as we
know, there is no theoretical upper bound on this constant --- it is
bounded only by current experimental constraints.  Applying these
constraints, we find that the prospects for signals at the Large
Hadron Collider (LHC) are truly spectacular --- the four-body final
states given above may have cross sections of picobarns or larger,
leading to obvious signals visible in the first year of the LHC.  We
also determine the predicted kinematic distributions from unparticles,
which are insensitive to overall rates.  These distributions provide
testable predictions that could be used to distinguish
multi-unparticle production from other possibilities for new physics.

\section{3-point Correlation Function}
\label{sec:3point}

We begin by assuming that the standard model is coupled to the
unparticle sector through a scalar operator $O$.  For a scalar,
unitarity requires dimension $d \ge 1$, but there is no upper
bound~\cite{Mack:1975je,Grinstein:2008qk}.  Motivated by prominent
supersymmetric examples~\cite{Intriligator:1995au}, we consider the
range $1 \le d < 2$. Modifications are required for vector and tensor
operators~\cite{Mack:1975je,Nakayama:2007qu,Grinstein:2008qk}, but we
expect that our primary conclusions apply to these cases as well.

Conformal invariance constrains the 3-point $O$ correlation function
to be
\begin{equation}
   \langle 0 |O (x)\, O (y)\, O^\dagger (0)|0 \rangle =
 \frac{C'_d}{|x - y|^d \, |x|^d \, |y|^d}\, ,
\end{equation}
where $C'_d$ is a constant determined by the unparticle self-coupling
strength.  In momentum space, it is then
\begin{eqnarray}
   \lefteqn{\langle 0|O (p_1)\, O (p_2)\, O^\dagger (p_1 + p_2) |0 \rangle
   = C'_d \int d^4 x\, d^4 y\, \frac{1}{|x - y|^d \,
   |x|^d\, |y|^d}\,
   e^{i p_1 \cdot x}\, e^{i p_2 \cdot y} \nonumber } \\
   &=& C'_d \int d^4 x\, d^4 y\, d^4 z\, \delta^4 [z - (x- y)] \,
   \frac{1}{|z|^d \, |x|^d \, |y|^d}\,  e^{i p_1 \cdot x}\,
   e^{i p_2 \cdot y} \nonumber \\
   &=& C'_d \int \frac{d^4 q}{(2 \pi)^4 }\, \int d^4 x\, d^4
   y\, d^4 z\,
   \frac{e^{i q \cdot [z - (x-y)]}\, e^{i p_1 \cdot x}\, e^{i p_2 \cdot y}}
   {|z|^d\, |x|^d\, |y|^d} \nonumber \\
   &=& C_d \int \frac{d^4 q}{(2 \pi)^4}\,
   [-q^2 - i \epsilon]^{\frac{d}{2} - 2}\,
   [-(p_1 - q)^2 - i \epsilon]^{\frac{d}{2} - 2}\,
   [-(p_2 + q)^2 - i \epsilon]^{\frac{d}{2} - 2}\, ,
\label{3point_correlation}
\end{eqnarray}
where the last equality makes use of the unparticle 2-point
correlation function~\cite{Georgi:2007si}.  $C_d$ is determined in
terms of $C'_d\, $; we choose to express our results in terms of
$C_d$.

\begin{figure}[tb]
\begin{center}
\includegraphics*[width=10cm,clip=]{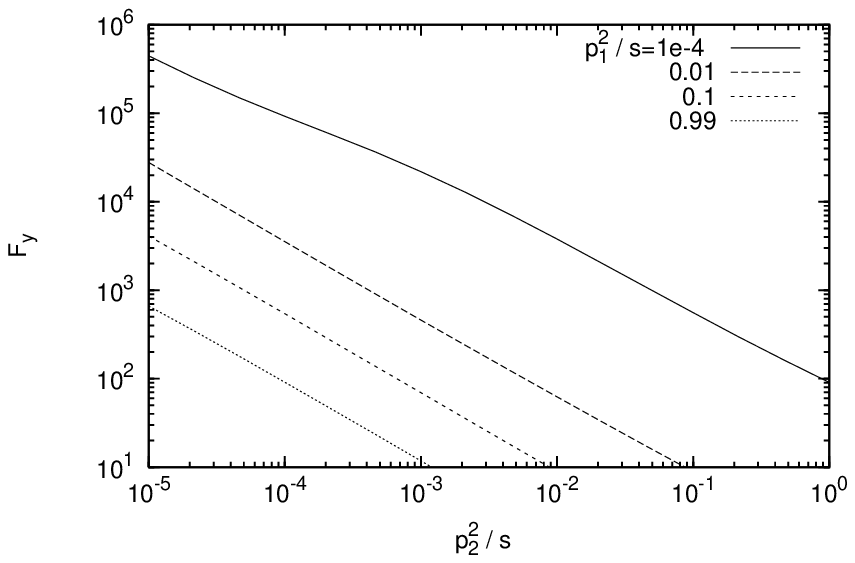}
\includegraphics*[width=10cm,clip=]{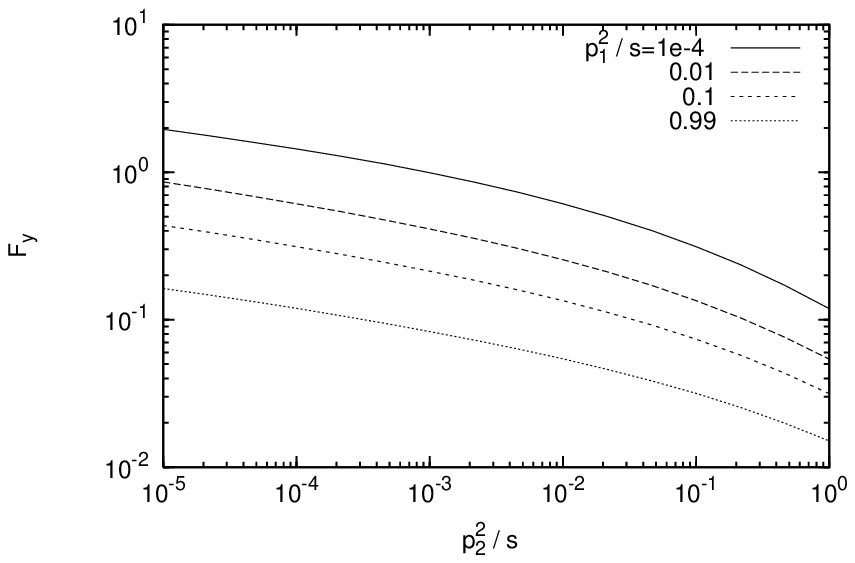}
\end{center}
\caption{Feynman parameter integral $F_y (p^2_1/s, p^2_2/s)$, as a
function of $p^2_2/s$ for several $p^2_1/s$ values and $d=1.1$ (top)
and 1.9 (bottom).
\label{fig:feynp}}
\end{figure}

The 3-point correlation function is the product of 3 propagators from
0 to $x$, $x$ to $y$, and $y$ to 0. It is therefore not surprising
that, in momentum space, it takes the form of a loop integral.  We may
therefore make use of the standard techniques available for
simplifying loop calculations.  Using Feynman parameters for
\eqref{3point_correlation}, the 3-point correlation function becomes
\begin{equation}
\langle 0|O (p_1) O (p_2) O^\dagger (p_1 + p_2) |0 \rangle =
-i\, (-1)^n \, C_d\, \left(\frac{1}{s}\right)^{n-2}
F_y \! \left(\frac{p^2_1}{s}, \frac{p^2_2}{s} \right) \, ,
\label{threepoint}
\end{equation}
where
\begin{equation}
F_y \! \left(\frac{p^2_1}{s}, \frac{p^2_2}{s} \right)
= \frac{1}{16 \pi^2} \frac{\Gamma (n-2)}{[\Gamma (\frac{n}{3})]^3}
 \int^1_0 d x_1 d x_2 d x_3
   \delta (x_1 + x_2 + x_3 -1)
   \left(\frac{1}{\Delta^\prime} \right)^{n-2}
   (x_1 x_2 x_3)^{1-\frac{d}{2}}\, ,
\end{equation}
$n = 6 - \frac{3 d}{2}$ and $\Delta^\prime = x_1 x_2\, p^2_1/ s + x_1
x_3 p^2_2 / s + x_2 x_3$, with $s = (p_1 + p_2)^2$.  $F_y (p^2_1/s,
p^2_2/s)$ is plotted in \figref{feynp}.

\section{Bounds on Unparticle Interactions}
\label{sec:interactions}

To investigate the phenomenological implications of the 3-point
correlation function discussed in \secref{3point}, we must introduce
unparticle couplings to standard model particles. Here we present here
the relevant couplings, and determine the existing constraints on them
and the prospects for probing them at the LHC.  The results here are
independent of unparticle self-interactions, but are of interest in
their own right.  In addition, in \secref{4gamma} we will present
results for a reference value of $\Lambda_4$, which we choose based on
the results we derive here.

A scalar unparticle may couple to two gauge bosons or two fermions
through the Lagrangian interaction terms
\begin{equation}
{\cal L} = \frac{c_i}{\Lambda_4^d} O F ^i _{\mu\nu} F^{ i \, \mu
  \nu} + \frac{e c_4 ^f}{\Lambda_4^d} O h \overline{f_L} f_R \ ,
\end{equation}
leading to the Feynman rules
\begin{eqnarray}
   O \gamma \gamma \, , \ Ogg \ \text{vertices}: &&
   i\, \frac{4 c_{\gamma, g}}{\Lambda^d_4}\,
   (-p_a \cdot p_b\, g^{\alpha \beta}
   + p^\beta_a\, p^\alpha_b) \nonumber \\
   O \bar{f} f \ \text{vertex}: &&
   i\, e\, \frac{c^f_4\, v}{\Lambda^d_4}\, P_R\, ,
\label{O_Feynmanrules}
\end{eqnarray}
where $\Lambda_4$ is some high scale characterizing these
non-renormalizable interactions, $v$ is the Higgs vacuum expectation
value, $e$ is the proton charge, and $c_{\gamma}$, $c_g$, and $c_4^f$
are constants.

Following convention, we choose an $O$ normalization by specifying the
unparticle propagator~\cite{Georgi:2007si,Cheung:2007zza}
\begin{equation}\label{unpartprop}
   {\rm scalar~unparticle~propagator}: \quad
   i\, B_d\, \theta (q^0)\, \theta (q^2)\, (q^2 - \mu^2)^{d-2}\, ,
\end{equation}
where
\begin{equation}
    B_d \equiv A_d\, \frac{\left(e^{-i \pi} \right)^{d-2}}
    {2 \sin d \pi}\, , \hspace{0.3cm}
   A_d \equiv \frac{16\, \pi^{5/2}\, \Gamma (d+ \frac{1}{2})}
    {(2 \pi)^{2d}\, \Gamma(d-1)\, \Gamma (2d)}\, .
\end{equation}
Here the modified propagator suggested in Ref.~\cite{Fox:2007sy} is
used to take into account the breaking of conformal invariance at a
scale $\mu$ by unparticle couplings to the Higgs boson.  A more
detailed analysis was performed in Ref.~\cite{Delgado:2007dx} by
considering a deconstructed version of the unparticle-Higgs
coupling. That approach led to a propagator of the form
$\left[(q^2)^{2-d}- v^2\, (\mu^2_U)^{2-d} / (q^2 - m^2_h)
\right]^{-1}$, where $v$ and $m_h$ are the Higgs vacuum expectation
value and mass, respectively, and $\mu_U$ is a scale related to the
Higgs-unparticle coupling.

We will be most interested in unparticle couplings to two photons.
This coupling, in conjunction with unparticle couplings to gluons and
quarks, mediates processes $gg , q \bar{q} \to \unp \to \gamma
\gamma$, which have been studied previously in
Ref.~\cite{Kumar:2007af}.  With our coupling conventions, these
processes have differential cross sections
\begin{eqnarray}
\overline{\left|{\cal M}_{gg \to 2 \gamma}\right|^2}
&=& 2\, \left|\frac{c_g\, c_{\gamma}}{\Lambda^{2d}_4}
\right|^2\, |B_d|^2\, (\hat{s} - \mu^2)^{2d-4}\, \hat{s}^4\, , \\
\overline{\left|{\cal M}_{q\bar{q} \to 2 \gamma}\right|^2}
&=& \frac{2}{3} \,
\left|\frac{e\, c^f_4\, c_{\gamma}}{\Lambda^{2d}_4}
\right|^2\, |B_d|^2\, v^2\, (\hat{s} - \mu^2)^{2d-4}\, \hat{s}^3\, ,
\end{eqnarray}
where we have averaged and summed over initial and final state colors
and polarizations, but have not yet included the factor 1/2 to account
for the two identical photons in the final state.

These processes are bounded by existing data, most stringently by the
diphoton invariant mass distribution from the CDF Collaboration at the
Tevatron~\cite{Aaltonen:2007kpa}, based on an integrated luminosity of
$1.2~\ifb$ at $\sqrt{s} = 1.96~\tev$.  The events in this distribution
have two central photons with $|\eta| < 1.04$ and transverse momenta
$p^{1,2}_T \geq 15~\gev$.  To obtain a lower bound on $\Lambda_4$, we
simulate the signal by adopting the following procedure here and in
all analyses described below. We set $c_g = c_\gamma = 1$ and $e^2\,
(c^f_4)^2 = 2 \pi$, following the convention justified in
Ref.~\cite{Bander:2007nd}, use the CTEQ5L parton distribution
functions with a factorization scale $\mu_f =
\sqrt{\hat{s}}$~\cite{Lai:1999wy}, and evaluate the cross section with
the Monte Carlo program VEGAS~\cite{Lepage:1977sw}.  We then impose
the identical $\eta$ and $p_T$ cuts given above, and further require
$m_{\gamma\gamma} > M_{\text{th}}$, where $m_{\gamma\gamma}$ is the
diphoton invariant mass, and $M_{\text{th}}$ is a threshold mass
chosen to maximize the sensitivity to the signal.  At the Tevatron,
unparticle production through $q\bar{q}$ dominates over $gg$, and the
most stringent bounds are typically achieved for $M_{\text{th}}
\approx 350~\gev$.  The 95\% CL lower bound on $\Lambda_4$ is then
derived following Ref.~\cite{Feldman:1997qc}; these results are given
in \figref{bounds}.  For comparison, in \figref{bounds} we also
present bounds on $\Lambda_4$ given previously in the literature from
$e^+ e^- \to \unp \to \mu^+ \mu^-$ from LEP/SLC~\cite{Bander:2007nd}
and the unitarity bound on $WW$ scattering~\cite{Greiner:2007hr}.

\begin{figure}[tb]
\begin{center}
\includegraphics*[width=10cm,clip=]{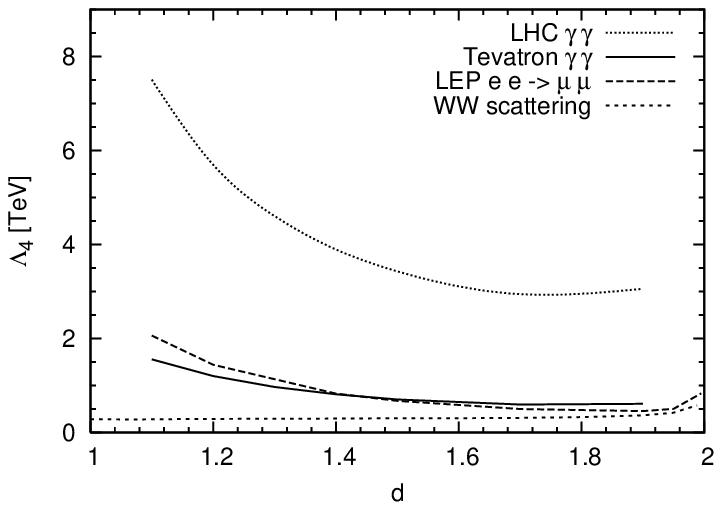}
\end{center}
\caption{Bounds on the unparticle scale $\Lambda_4$ from $p \bar{p}
\to \unp \to \gamma \gamma$ using existing data from the Tevatron
(solid), and the projected $5\sigma$ discovery reach at the LHC for
integrated luminosity $100~\ifb$ (dotted).  For comparison, existing
bounds from $e^+ e^- \to \unp \to \mu^+ \mu^-$~\cite{Bander:2007nd}
(long dashed) and the unitarity bound on $WW$
scattering~\cite{Greiner:2007hr} (short dashed) are also shown.  See
text for details. }
\label{fig:bounds}
\end{figure}

We have also determined the discovery reach for $p p \to \unp \to
\gamma \gamma$ at the LHC, assuming $100~\ifb$ of data.  We evaluate
the prompt diphoton background using PYTHIA
6.4~\cite{Sjostrand:2006za}, impose kinematic cuts $|\eta| < 2.5$ and
$p_T^{1,2} > 25~\gev$, and again optimize by varying the threshold
energy $M_{\text{th}}$. We estimate a $5\sigma$ discovery reach by
requiring $S/ \sqrt{B} > 5$ and $S > 5$, where $S$ and $B$ are the
number of signal and background events.  This discovery reach is also
presented in \figref{bounds}.

These results have omitted important effects.  Notably, in determining
the LHC reach, we have omitted the background from misidentification
of jets and electrons.  Nevertheless, we may conclude that present
limits from the Tevatron and LEP/SLC are comparable, and imply lower
bounds on $\Lambda_4$ of $\sim 2~\tev$ for $d=1.1$, dropping to below
a TeV for $d=1.9$.  We will adopt the reference value $\Lambda_4 =
1~\tev$ below.  At the LHC, the sensitivity is much higher, with the
$\gamma \gamma$ signal probing $\Lambda_4$ in the multi-TeV range.
Note that here, and in the rest of this work, we have taken the
conformal breaking scale to be $\mu = 0$ in deriving our results.
However, we have found that our results are insensitive to this
choice, provided $\mu \alt M_{\text{th}}$.  In general, the
modifications have the property that they are small for large
$q^2$. For example, for $q^2 \gg \mu^2$, the propagator we used
(Eq.~(\ref{unpartprop})) is of the form $(q^2)^{d-2}$ with corrections
of order $\mu^2 / q^2$.  In this limit the propagator obtained from
Ref.~\cite{Delgado:2007dx} has corrections of order $(\mu^2 /
q^2)^{2-d}\, v^2 / (q^2 - m^2_h)$.  We will assume henceforth that the
breaking of conformal symmetry is small in the kinematic region we
work in so that the correction terms can be ignored.

\section{Four Photon Events}
\label{sec:4gamma}

We now turn to new processes mediated by unparticle
self-interactions. As noted in \secref{intro}, the 3-point function is
particularly interesting, and, depending on the unparticle
interactions with the standard model, may mediate a variety of
processes leading to spectacular 4-body final states.  As an example,
in this section we consider the four photon signal shown in
\figref{4gamma_feynmandiagram}.

The cross sections for $g g, q \bar{q} \to \unp \to \unp \unp \to
\gamma \gamma \gamma \gamma$ are completely specified, given the
Feynman rules of \eqsref{threepoint}{O_Feynmanrules}.  With momenta as
labeled in \figref{4gamma_feynmandiagram}, we find
\begin{eqnarray}
\overline{\left|{\cal M}_{gg \to 4 \gamma}\right|^2}
&=& 2^{10}\,
\left|\frac{c_g \, c_{\gamma}^2}{\Lambda^{3d}_4} \right|^2 \,
(p_a \cdot p_b)^2 \, (p_1 \cdot p_2)^2 \, (p_3 \cdot p_4)^2 \,
|\langle 0|O (q_2) O (q_3) O^\dagger (q_1) |0 \rangle|^2 \\
\overline{\left|{\cal M}_{q\bar{q} \to 4 \gamma}\right|^2}
&=& \frac{2^9}{3} \,
\left| \frac{e \, c^f_4 \, c_{\gamma}^2}{\Lambda^{3d}_4} \right|^2 \,
v^2 (p_a \cdot p_b) \, (p_1 \cdot p_2)^2 \, (p_3 \cdot p_4)^2 \,
   |\langle 0|O (q_2) O (q_3) O^\dagger (q_1) |0 \rangle|^2 \, ,
\end{eqnarray}
where we have averaged and summed over initial and final state colors
and polarizations, but have not yet included the factor 1/4!~to
account for the four identical photons in the final state.  To
evaluate the parton-level cross sections, we use the results of
\secref{3point} for the 3-point correlation function.  Setting $c_g =
c_\gamma = 1$ and $e^2\, (c^f_4)^2 = 2 \pi$ as above, and integrating
over 4-body phase space with the Monte Carlo program
VEGAS~\cite{Lepage:1977sw}, we find that the parton-level cross
sections may be written as
\begin{eqnarray}
    \hat{\sigma}_{g g \to 4 \gamma} (\hat{s})
    &=& f^g_d\, C_d ^2 \left(\frac{\hat{s}}{\Lambda^2_4} \right)^{3 d}\,
    \frac{1}{\left(\hat{s} / [{\rm GeV}^2] \right)}~[\fb] \\
\label{pcs_qqbar}
    \hat{\sigma}_{q \bar{q} \to 4 \gamma} (\hat{s})
    &=& f^q_d\,  C_d ^2 \left(\frac{\hat{s}}{\Lambda^2_4} \right)^{3 d}\,
    \left(\frac{v^2}{\hat{s}} \right)\,
\frac{1}{\left(\hat{s} / [{\rm GeV}^2]
    \right)} ~[\fb] \, ,
\end{eqnarray}
where the dimensionless proportionality factors $f^g_d$ and $f^q_d$
are given in \tableref{Total_cs}.

\begin{table}[t!]
\begin{center}
\begin{tabular}{|c|c|c|c|c|}
\hline
& d=1.1 & d=1.2 & d=1.5 & d=1.9 \\
\hline
\hline
$f^g_d$ & 2.7 & 1.2 & 0.16 & 0.02  \\
\hline
$f^q_d$ & 5.5 & 2.5 & 0.35 & 0.04  \\
\hline
Tevatron $\sigma_{q \bar{q} \to 4 \gamma}^{\text{ref}}$ [fb]
& $2 \times 10^{-8}$ & $6 \times 10^{-9}$ & $2.5 \times 10^{-10}$
& $1.5 \times 10^{-11}$ \\
\hline
Tevatron $\sigma_{g g \to 4 \gamma}^{\text{ref}}$ [fb] &
$3 \times 10^{-9}$ &
$7 \times 10^{-10}$ & $2 \times 10^{-11}$ & $6 \times 10^{-13}$ \\
\hline
Tevatron upper bounds on $C_d/(\Lambda_4~{\rm [TeV]})^{3d}$ &
$1.3 \times 10^4$ & $2.3 \times 10^4$ &
$1.2 \times 10^5$ & $4.8 \times 10^5$ \\
\hline
\hline
LHC $\sigma_{g g \to 4 \gamma}^{\text{ref}}$ [fb] &
$2.6 \times 10^{-5}$ & $1.7 \times 10^{-5}$ & $1.3 \times 10^{-5}$
& $3.5 \times 10^{-5}$ \\
\hline
LHC $\sigma_{q \bar{q} \to 4 \gamma}^{\text{ref}}$ [fb]
& $1.0 \times 10^{-6}$ & $6 \times 10^{-7}$ &
$3.6 \times 10^{-7}$ & $1.0 \times 10^{-6}$ \\
\hline
LHC maximum cross section [fb] & 4300 & 9600 & $1.8 \times 10^5$
& $8.4 \times 10^6$ \\
\hline
\hline
ILC $\sigma_{e^+ e^- \to 4 \gamma}^{\text{ref}}$ [fb] &
$1.0 \times 10^{-6}$ & $4.7 \times 10^{-7}$ & $6.0 \times 10^{-8}$ &
$8.0 \times 10^{-9}$ \\
\hline
ILC maximum cross section [fb] & 160 & 250 & 810 & 1900 \\
\hline
\end{tabular}
\end{center}
\caption{Dimensionless parton-level proportionality factors
$f^{g,q}_d$, and unparticle $4\gamma$ reference cross sections at the
Tevatron ($\sqrt{s} = 1.96~\tev$), LHC ($\sqrt{s} = 14~\tev$) and ILC
($\sqrt{s} = 1~\tev$).  These reference cross sections assume
$\Lambda_4 =1~\tev$ and $C_d = 1$; the actual cross sections scale
with $C_d^ 2 /\Lambda_d ^ {6d}$.  Upper bounds on this combination of
parameters from existing data at the Tevatron are also listed, as are
the resulting maximum cross sections possible at the LHC and ILC. }
\label{table:Total_cs}
\end{table}

\subsection{Bounds from Tevatron}

The $4\gamma$ unparticle signal is bounded by searches at the
Tevatron.  The D0 collaboration has searched for the inclusive
production of multi-photon final states~\cite{Dzero:5067}.  Events are
selected with three or more photons in the central calorimeter
($|\eta| < 1.1$) and $E_T$-ordered cuts on their transverse energies:
$E^{1,2,3\, (,4)}_T > 30, 20, 15 \, (,15)~\gev$.  The dominant
backgrounds are diphoton production with additional initial state
radiation (ISR) photons and events in which jets or electrons are
misidentified as photons.  No excess of events above the standard
model prediction was found in the Tevatron data with integrated
luminosity $0.83 \pm 0.05$ fb$^{-1}$ collected during 2002--2005.

This search result thus sets upper bounds on combinations of the
unparticle self-interaction strength $C_d$ and the energy scale
$\Lambda_4$. To derive these bounds, we again set $c_g = c_\gamma = 1$
and $e^2\, (c^f_4)^2 = 2 \pi$.  As with the $\gamma \gamma$ signal
discussed in \secref{interactions}, at the Tevatron, the $q \bar{q}$
contribution dominates over that from $gg$, since $p \bar{p}$
collisions provide a large density of anti-quarks, and at $\sqrt{s} =
1.96~\tev$, large Bjorken $x$ is required.

We impose the kinematic cuts on rapidities and transverse energy as
given above.  Estimating the background event rate to be ${\cal L}\,
\sigma^{\rm SM}_{q \bar{q} \to 4 \gamma} \approx {\cal L}\, (\sum_q
Q_q^2)\, (\alpha / \pi)^2\, \sigma^{\rm SM}_{q \bar{q} \to 2 \gamma}
\sim 0.83~{\rm fb}^{-1}\, \times (0.002)^2 \times 0.66~{\rm pb} \sim
{\cal O} (10^{-3})$, we find that it is negligible.  The 95\% CL upper
limit of 3.04 events for zero background and zero events
observed therefore becomes the 95\% CL bound
\begin{equation}
   C^2_d\, \left(\frac{1}{\Lambda_4~[{\rm TeV}]} \right)^{6 d} \leq
   \frac{3.04}{0.83~{\rm fb}^{-1}\, \sigma^{\text{ref}}_{\text{tot}}}\, ,
\label{Tevatronbound}
\end{equation}
where $\sigma^{\text{ref}}_{\text{tot}} \equiv \sigma_{q \bar{q} \to 4
\gamma} ^{\text{ref}} + \sigma_{g g \to 4 \gamma} ^{\text{ref}}$ is
the total reference cross section, determined by setting $\Lambda_4 =
1~\tev$ and $C_d = 1$.  The resulting upper bounds on $C_d
/\Lambda_4^{3d}$ are given in \tableref{Total_cs}.

\subsection{Prospects for LHC and ILC}

We now determine the LHC cross sections for $4\gamma$ production
mediated by unparticle self-interactions.  Using
VEGAS~\cite{Lepage:1977sw}, we Monte Carlo simulate 4 to 10 million
events for several values of $d$.  We require the photons to have
rapidity $|\eta| < 2.5$ and transverse energies $E^{1,2,3,4}_T > 30,
20, 15, 15~\gev$.  The resulting reference cross sections for
$\Lambda_4 = 1~\tev$ and $C_d = 1$ from $gg$ and $q\bar{q}$ initial
states are given in \tableref{Total_cs}.

Of course, the actual cross sections scale with $C_d^ 2$.  We do not
know of any way to bound $C_d$ theoretically, or, indeed, to specify a
``typical'' value for $C_d$.  Lacking theoretical guidance, we simply
impose existing experimental bounds.  Although bounds on unparticles
have been discussed at length in the literature, these constraints,
including those derived and presented in \secref{interactions}, do not
bound $C_d$.  For a direct bound, we must therefore turn to the
Tevatron results derived above, which, of course, constrain $C_d$ and
$\Lambda_4$ in the combination identical to that which enters the LHC
cross section.  Assuming these parameters saturate the bound of
\eqref{Tevatronbound}, we find the maximal cross sections presented in
\tableref{Total_cs} and plotted in \figref{max_cs}.  These cross
sections are extraordinarily large, ranging from picobarns for $d=1.1$
to nanobarns for $d=1.9$.  If the unparticle self-interaction is
anywhere near the largest values allowed by current experimental
constraints, this spectacular signal will be discovered very early at
the LHC.

\begin{figure}[tb]
\begin{center}
\includegraphics*[width=10cm,clip=]{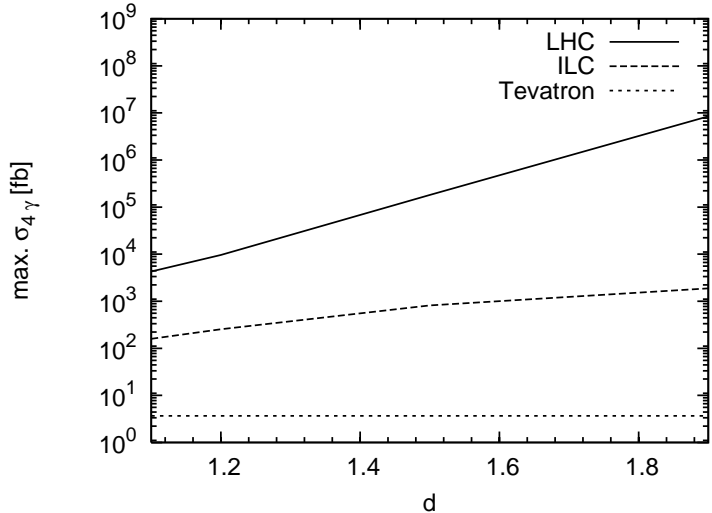}
\end{center}
\caption{Maximum possible cross sections for $4\gamma$ production
through unparticle self-interactions, given existing bounds on
$4\gamma$ events from the D0 Collaboration at the
Tevatron~\cite{Dzero:5067}.
\label{fig:max_cs}}
\end{figure}

The $p_T$ distributions of the leading, next-to-leading,
next-next-to-leading, and fourth photons in unparticle $4\gamma$
events are given in \figref{pt_spec}.  We find that the $p_T$ spectra,
even for the 3rd and 4th photon, are remarkably hard.  Given this, the
standard model background of diphoton production with two ISR photons
can be eliminated with $p_T$ cuts with little effect on the signal.
The dominant backgrounds will be from events with misidentified jets
or electrons, but even these may be reduced significantly with hard
$p_T$ cuts without large reduction in the signal.  It would be very
interesting to determine the extent to which this is validated by a
realistic detector study.  Last, we note that the shapes of the $p_T$
distributions, along with other kinematic information, are, of course,
independent of $C_d$ and $\Lambda_4$.  They may therefore be used to
identify unparticles as the source of these signals, and to
distinguish them from anomalies predicted by other frameworks for new
physics.

\begin{figure}[tb]
\begin{center}
\includegraphics*[width=10cm,clip=]{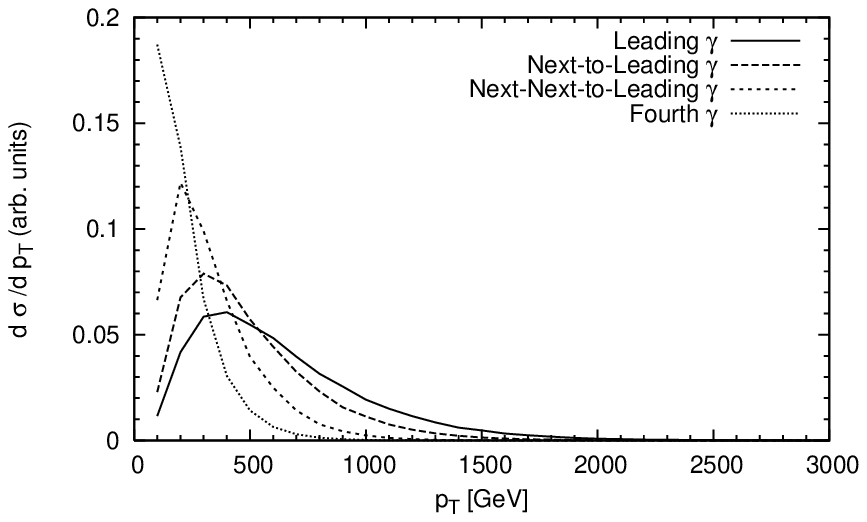}
\includegraphics*[width=10cm,clip=]{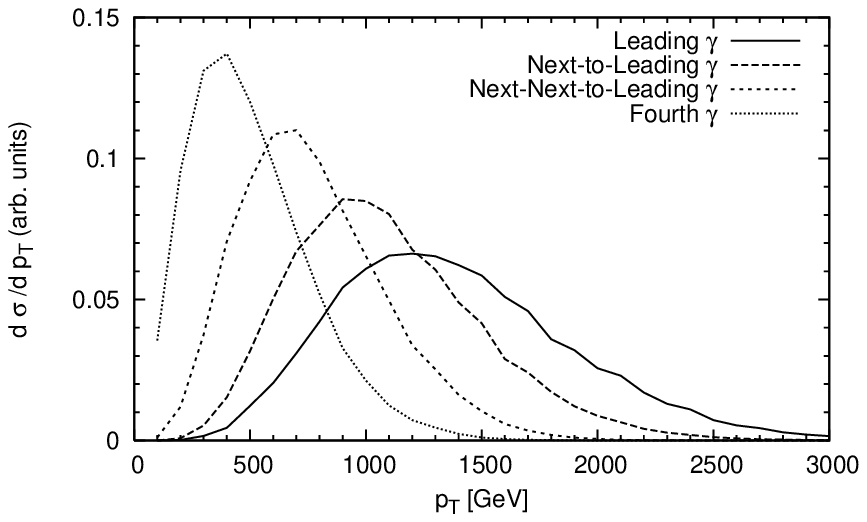}
\end{center}
\caption{The $p_T$ distributions for $p_T$-ordered photons in
unparticle $4\gamma$ events (from the dominant $gg$ initial state
only) at the LHC for $d=1.1$ (top) and 1.9 (bottom).
\label{fig:pt_spec}}
\end{figure}

If the coupling of scalar unparticles to electrons is of the same size
as to quarks, $4\gamma$ event rates at the International Linear
Collider (ILC) can be calculated with \eqref{pcs_qqbar} multiplied by
3.  ILC reference cross sections and maximal possible cross sections
are presented in \tableref{Total_cs} and \figref{max_cs}.
Interestingly, the $4\gamma$ cross section at a 1 TeV ILC is about two
orders of magnitude larger than at the Tevatron.  This is because the
parton densities become less than one for $x \agt 0.3$, so the
Tevatron $p \bar{p} \to 4 \gamma$ cross section is dominated by the
contribution from $\sqrt{\hat{s}} \sim 300~\gev$.

\section{Discussion}

In this work, we have investigated a new feature of unparticle
physics, namely, the self-interactions of unparticles.  Such
interactions are necessarily present in conformal theories, are
unsuppressed for strongly coupled conformal sectors, and introduce a
large range of new phenomena not studied previously.

As an example, we have investigated the 3-point correlation function
in detail. This is completely specified by conformal
invariance,\footnote{The importance of self-interactions in the case
of broken conformal symmetry has also been recently emphasized in
Ref.~\cite{Strassler:2008bv}.} up to a constant $C_d$, and mediates a
host of processes, such as $pp \to \unp \to \unp \unp \to \gamma
\gamma\gamma\gamma, \gamma\gamma ZZ, ZZZZ, \gamma\gamma l^+ l^-, ZZl^+
l^-, 4l$ at the LHC.  We know of no upper bound on $C_d$.  We have
therefore imposed only the existing Tevatron bounds, and have found
that, for the example $4\gamma$ signal discussed here, the allowed
cross sections at the LHC are enormous, ranging from picobarns for
unparticle dimension $d=1.1$ to nanobarns for $d=1.9$.  Such signals
could emerge very early after the LHC turns on, and imply that the
4-body final states could be by far the most promising modes for
unparticle discovery at colliders.  Given these observations, it would
be very interesting to perform similar analyses for the other final
states, as well as to carry out realistic detector simulations to
determine the extent to which jet and lepton misidentification masks
these {\em a priori} spectacular signals.

On the theoretical side, the most pressing issue is to determine what
values of $C_d$ are possible or natural.  Conformal invariance by
itself does not determine the constant, and we are unaware of any
other consistency conditions that could bound $C_d$.  It would be very
interesting to see if the AdS/CFT correspondence can shed light on
this issue. In the absence of theoretical bounds, $C_d$ can be large,
and $p p \to \unp \unp \to 4\gamma$ and related 4-body final states
may have cross sections larger than $p p \to \unp \to \gamma\gamma$
and related 2-body states.  Such a conclusion is highly
counter-intuitive, but perhaps possible, given the other surprising
properties of unparticles discovered so far.

\begin{acknowledgments}
We thank Yuri Shirman and Matt Strassler for interesting
discussions. The work of JLF is supported in part by NSF Grant
Nos.~PHY--0239817 and PHY--0653656, NASA Grant No.~NNG05GG44G, and the
Alfred P.~Sloan Foundation. The work of AR is supported in part by NSF
Grant Nos.~PHY--0354993 and PHY--0653656.  The work of HT is supported
in part by NASA Grant No.~NNG05GG44G.
\end{acknowledgments}

\end{document}